\documentclass[a4paper,11pt]{article}
\pdfoutput=1 

\usepackage{jcappub} 

\usepackage[T1]{fontenc} 
\usepackage{multicol}        
\usepackage{multirow}        
\usepackage{tabu}      

\title{Oscillation modes of hybrid stars within the relativistic Cowling approximation}
\author[a,b]{Ignacio F. Ranea-Sandoval}
\author[c,d]{Octavio M. Guilera}
\author[a,b]{Mauro Mariani}
\author[a,b]{Milva G. Orsaria}

\affiliation[a]{Grupo de Gravitaci\'on, Astrof\'isica y Cosmolog\'ia, Facultad de Ciencias Astron{\'o}micas y Geof{\'i}sicas, Universidad Nacional de La
  Plata, Paseo del Bosque S/N (1900), La Plata, Argentina.}
\affiliation[b]{CONICET, Godoy Cruz 2290, 1425, Buenos Aires, Argentina}
\affiliation[c]{Instituto de Astrof\'isica de La Plata, CONICET, Argentina}
\affiliation[d]{Grupo de Ciencias Planetarias, Facultad de Ciencias Astron\'omicas y Geof\'isicas, Universidad Nacional de La Plata, Paseo del Bosque S/N (1900), La Plata, Argentina}

\emailAdd{iranea@fcaglp.unlp.edu.ar}
\emailAdd{oguilera@fcaglp.unlp.edu.ar}
\emailAdd{mmariani@fcaglp.unlp.edu.ar}
\emailAdd{morsaria@fcaglp.unlp.edu.ar}

\abstract{The first direct detection of gravitational waves has opened a new window to study the Universe and would probably start a new era: the gravitational wave Astronomy. Gravitational waves emitted by compact objects like neutron stars could provide significant information about their structure, composition and evolution.
  
In this paper we calculate, using the relativistic Cowling approximation, the oscillations of compact stars focusing on hybrid stars, with and without a mixed phase in their cores. We study the existence of a possible hadron-quark phase transition in the central regions of neutron stars and the changes it produces on the gravitational modes frequencies emitted by these stars. We pay particular attention to the $g$-modes, which are extremely important as they could signal the existence of pure quark matter inside neutron stars. Our results show a relationship between the frequency of the $g$-modes and the constant speed of sound parametrization for the quark matter phase. We also show that the inclusion of color superconductivity produces an increase on the oscillation frequencies.

We propose that observations of $g$-modes with frequencies $f_{\rm g}$ between $1$ kHz and $1.5$ kHz should be interpreted as an evidence of a sharp hadron-quark phase transition in the core of a compact object.}

\begin{document}
\maketitle
\flushbottom

\section{Introduction}
\label{sec:intro}
The discovery of the $2M_\odot$ pulsars PSR J1614-2230 \cite{2solar1} and PSR J0348-0432 \cite{2solar2} has imposed restrictions to the equation of 
state (EoS) needed to describe matter inside compact objects \cite{lattimerscience,Lattimer:2010uk}. The precise determination of their 
masses has forced astrophysics to rethink the internal composition of neutron stars (NSs).

Matter in the inner core of NSs is compressed to densities several times higher than the density of an ordinary atomic nuclei. At such densities, atomic nuclei are squeezed so tightly together that a phase transition to a free quark state may occur. To describe matter in the central regions of NSs different hybrid EoSes with exotic matter (considering hyperonic, quark and/or diquark degrees of freedom) have been proposed \cite{Bona:2012,Orsaria:2013,Orsaria:2014}. {{Other theoretical possibilities such as appearance of mesonic condensate (Pions and/or Kaons) produce EoS too soft to construct stars compatible with the $2M_\odot$ observations \citep[see, for example,][]{Kaon2001LNP,Kaon2006ARNPS,Kaon2010arXiv1012.3208L,Kaon2012PhRvD}. Moreover, the inclusion of mesonic condensates produce strong changes in the cooling rates of neutron stars and the numerical results are not in agreement with the observational cooling curves of compact objects \cite[see, for example,][]{cooling2004ARA&A}.}}

The phase transition that might occur between the inner and outer core of a NS could be described within different scenarios, depending on the value of the hadron-quark surface tension, $\sigma _{\rm HQ}$. There are two limiting cases, the Maxwell construction, which assumes that $\sigma _{\rm HQ}$ is infinitely large and the ``bulk'' Gibbs construction, which assumes that $\sigma _{\rm HQ}$ is zero. For intermediate values of $\sigma _{\rm HQ}$ , the ``full'' Gibbs formalism must be used \cite{finite-size-WU_SHEN}. Within this scenario, the detailed geometrical structures that appear in the pasta-phase can be studied and a relationship with $\sigma _{\rm HQ}$ might be possible to obtain. Due to the uncertainties in the hadron-quark surface tension value, the nature of the suggested hadron-quark phase transition is not clear. Theoretical studies suggests that, if the surface tension is grater than a critical value $\sim 5 - 40\,{\rm MeV/fm}^2$, the phase transition is sharp and the Maxwell construction would be favored \citep{Alford:2001zr,endo:2011}. Otherwise, a mixed phase exists and the phase transition should be described through the Gibbs construction. There are theoretical results that show, for some EoSes, that larger values of $\sigma _{\rm HQ}$ shorten the extension of the mixed phase \cite{fullGibbs:2006,fullGibbs:2014}. Despite this fact, a correlation between these quantities is not clear because the results are EoS dependent.

From here on, NSs with pure quark matter in their cores will be referred as quark-hybrid stars to distinguish them from the hybrid stars in which a mixed phase of quark and hadrons is present.

\begin{figure*}[h]
  \centering
  \includegraphics[width=0.4\textwidth]{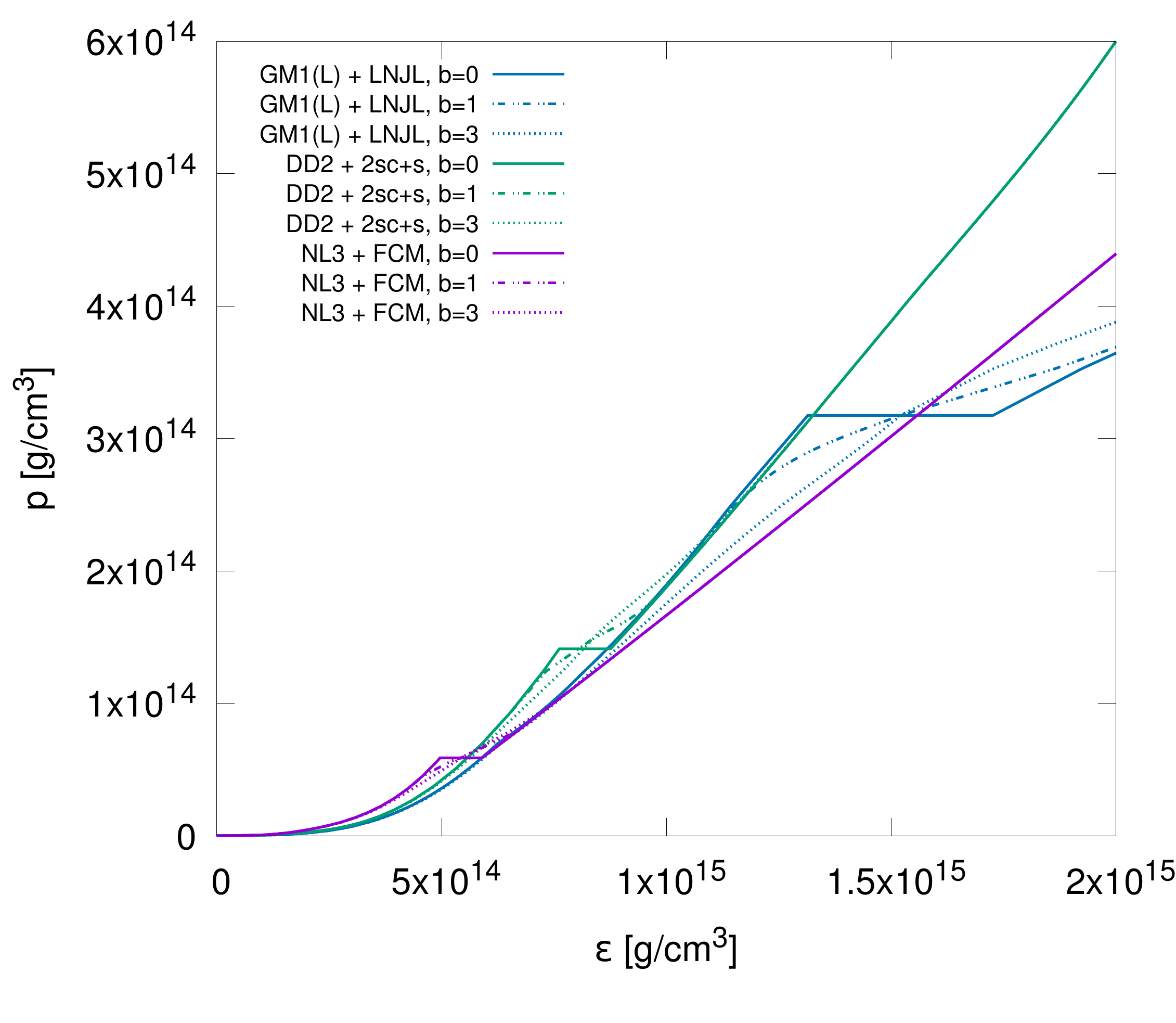}\includegraphics[width=0.4\textwidth]{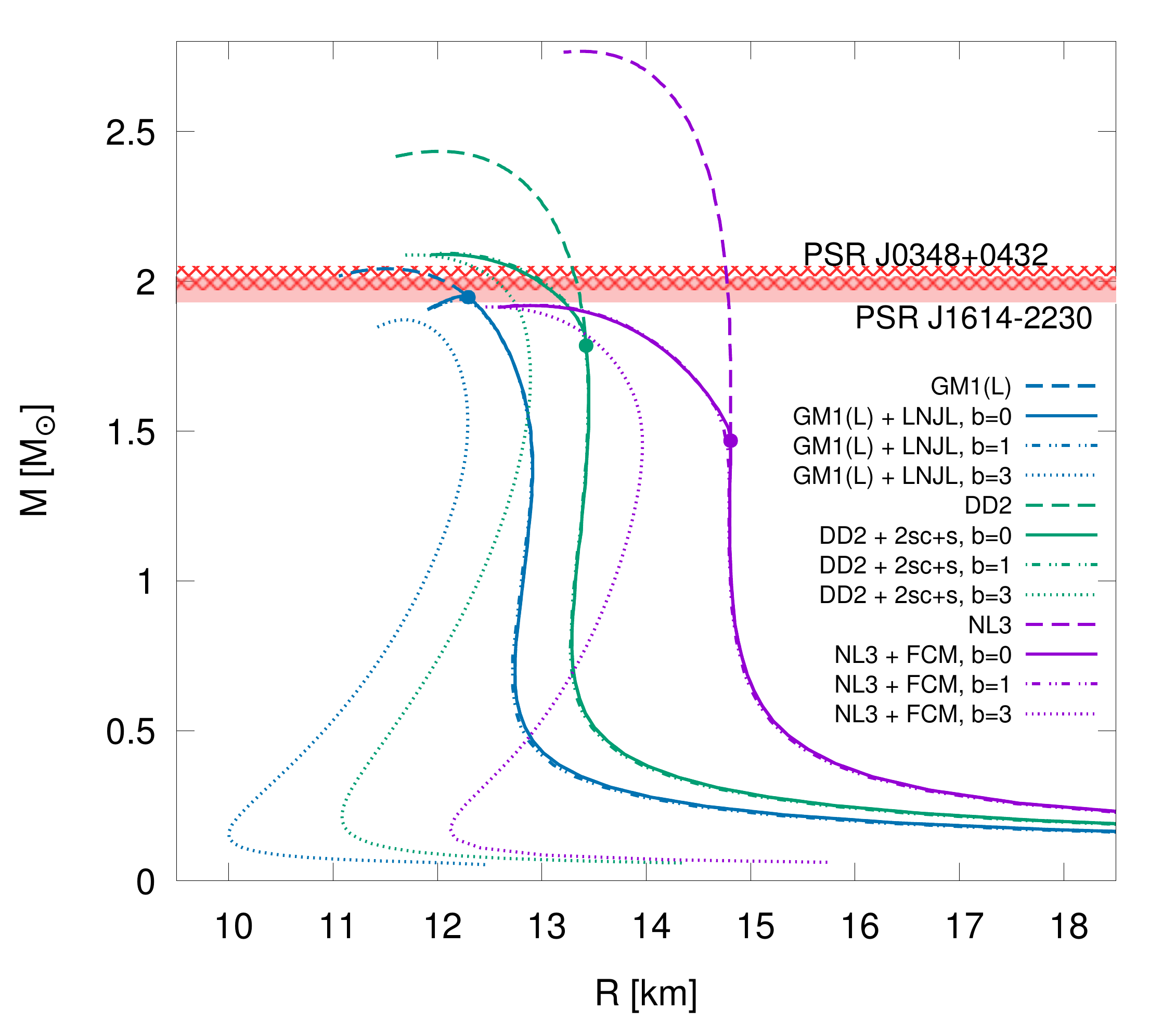}
  \caption{(Color online) Left panel: A sample of the hybrid EoSes used in this work. The results are obtained using the Maxwell formalism (solid lines) and mixed phase to construct the phase transition (dot dashed and dotted lines). Right panel: Using the same color and line type convention we present the corresponding mass-radius relationships and the measured masses (and error bars) of PSR J1614-2230 and PSR J0348-0432. In the case of sharp phase transitions, the dot in the M-R curves represents the first star with a pure quark matter core. {{Purely hadronic families of stationary stellar models are shown, for completeness, with dashed lines.}}
}
  \label{eos-mraio}
\end{figure*}

The structure of NSs depends mainly on the EoS used to describe their internal composition. Stationary stellar configurations have been studied since the pioneering works of Tolman and Oppenheimer and Volkoff, in which the fundamental equations of relativistic hydrostatic equilibrium were coined: the TOV equations \cite{tolman39:a,oppenheimer39:a}. Given a particular EoS, these equations allow to obtain the corresponding family of stationary stellar configurations. These theoretical curves could be compared with the observed masses and radii of compact objects to discard those unable to reproduce astronomical observations \cite{Ozel:2016}.

Oscillation modes are known to be extremely sensitive to internal composition of compact stars. In particular, the gravity modes are a useful tool to probe the composition of NSs and to test their internal structure. Moreover, they could provide significant information about the nature of the hadron-quark phase transition  and the existence (or the absence) of deconfined quark matter in their inner cores \cite{AndersonKokkotas:1996,AndersonKokkotas:1998,GWfromNS,vasquez-lugones:2013}. 

Several oscillation modes in NSs may emit gravitational waves (GWs). The study of non-radial pulsations of NSs has started at the end of the 60s \cite{NS-oscillations}. Later, further works about GW asteroseismology in NSs have been developed \cite{GWfromNS}. Following the mode classification introduced by Cowling, the study of stellar oscillation dumped by GW emission (quasi-normal modes) could be used as a tool to shed some light on the nature of matter inside such objects \cite{cowling:1941}.

After decades of development, LIGO and VIRGO collaborations have detected directly GWs emitted from binary black hole mergers: GW150914 \cite{GW1}, GW151226 \cite{GW2}, GW170114 \cite{GW3} and GW170814 \cite{GW4} and from a merger of two neutron stars, GW170817 \cite{GW5} whose electromagnetic counterpart (GRB 170817A) has also been observed \cite{GW-EM}. These observations confirm the last prediction of Einstein's General Relativity theory and marks the beginning of a new era in Astronomy: the GW window has been opened.

In this work we will study pulsation modes of compact stars using the relativistic Cowling approximation \cite{sotani-cowling}. We will focus on EoSes for which the stationary family of stars have a maximum mass compatible with the observations of PSR J1614-2230, $M=(1.928 \pm 0.017)M_\odot$, and 
PSR J0348-0432, $M=(2.01 \pm 0.04)M_\odot$. We will also pay particular attention to hybrid EoSes with quark matter and their corresponding family of quark-hybrid stars.

The paper is organized as follows. In Section \ref{EoSes} we will briefly present the EoSes used in this work and discuss some of their properties. 
Section \ref{cowling} will be mainly devoted to present the relativistic Cowling approximation, its range of applicability and the comparison with other methods to compute stellar oscillations of compact objects. Besides, we will present the numerical method used and the results obtained through the {\it Cowling Frequency Key} (CFK) code developed for this work. In Section \ref{results} we will show the main results of the paper. The summary, discussion and the astrophysical implications of these results will be provided in Section \ref{discussion}. 

\section{The equations of state} \label{EoSes}

The structure of NSs can be described by three main layers: the crust, the outer core and the inner core. Matter in each of these layers has extremely different characteristics. For this reason, it is usually described using different models. For the crust, which is formed by a crystalline lattice of ions, we use a combination of the Baym-Bethe-Pethick and Baym-Pethick-Sutherland EoS \cite{Baym:1971,BAYM2}. To describe matter in the outer core, which is composed of hadronic matter, we use several models that will be described below. The nature of matter in the inner core is not completely understood, but it is suggested that a phase of deconfined quark matter could exist. Thus, to describe the inner core we use different quark EoSes. The hadron-quark phase transition is modeled through two different approaches described in this Section.

\subsection{Hadronic matter}

To describe matter in the outer core of hybrid stars we use the 
non-linear relativistic mean field theory
\cite{walecka:1974,boguta:1977,boguta:1983}. Within this theory, it is assumed that interactions between baryons (neutrons, protons, hyperons) and delta resonances are mediated by scalar, $\sigma $, vector, $\omega$, and isovector, $\rho $, mesons fields. 
The mean-field equations are obtained after replacing the operators of meson fields by their respective ground-state expectation values. The coupling constants for the different types of interactions are given by $g_{\sigma B}=x_{\sigma B}g_{\sigma}$, $g_{\omega B}=x_{\omega B}g_{\omega}$, and $g_{\rho B}=x_{\rho B}g_{\rho}$. We consider $x_{\sigma B}=0{.}7$ and $x_{\omega B}=x_{\rho B}=1$ for the NL3 parametrization \cite{rs-etal:2016}. {The analysis of the observational data from event GW170817 constraint the tidal deformability parameter for a compact object with a mass of $1.4 M_\odot$ to be $\Lambda_{1.4} < 800$  with a confidence level of $90\%$ \cite{PhysRevLett.121.161101}. This limiting value and the results obtained using several realistic EOSes as well as EOSes that interpolate between theoretical results at low and high baryon densities, constraint the radius of a compact object with $M = 1.4M_\odot$ to be $R_{1.4} \lesssim 13.6$km (see, for example, \cite{PhysRevLett.120.172703} and references therein). This restriction discards the NL3 parametrization to describe hadronic matter. However, we use NL3 parametrization to compare our results with previous works and gain confidence about the robustness of the numerical code we developed (see, for example, \cite{vasquez-lugones:2013}). In Subsection \ref{numerical} below, we give a more detailed discussion about the comparison of our calculations with previous works.}

We have also used the DD2 parametrization with density-dependent coupling constants \cite{Typel:2005,typel2010}. Finally, a modification of the GM1 parametrization \cite{Glendenning:1985} is considered. In this modified EoS, called GM1(L), a density-dependent isovector meson-baryon coupling constant is included. This feature allows to fix (via an extra parameter) the slope of the asymmetry energy at nuclear saturation density \cite{Spinella:2017,RS-prep-2sc}.

\subsection{Quark matter}

In this work, we use several EoSes to describe quark matter in the inner core of hybrid stars. One of the models considered is the SU(3) local NJL (LNJL) model with vector interactions in which the only free parameter is the vector interaction coupling, $G_{\rm v}$, used to characterize the corresponding EoS \cite{Orsaria:2013,Orsaria:2014}. We also use the SU(3) non-local NJL (NLNJL) \cite{Orsaria:2014}. One of the main differences between LNJL and NLNJL is the inclusion of a Gaussian form factor in the later, which leads to momentum-dependent constituent quark masses. Like in the local model, vector interactions are included, being the vector coupling constant the free parameter used to characterize the EoS. The sets of parameters used for both the LNJL and NLNJL models are listed in \cite{Orsaria:2013,Orsaria:2014}.

In addition, we consider a model in which color superconductivity in the 2SC+s phase is included within the LNJL model \cite{RS-prep-2sc}. This model has three free parameters: the vector coupling constant, $\eta_v$, the diquark coupling constant, $\eta_{qq}$ and the effective strange quark mass in vacuum, $M_s^0$. These parameters are used to characterize the EoS.

Finally, we use zero-temperature limit of the Field Correlator Method (FCM) \cite{Dosch:1987-FCM,Simonov:1988-FCM,FCM-b,FCM-a,extendedFCM:2009}. FCM method is a non-perturbative approach to QCD that has been tested showing good agreement with Lattice-QCD calculations. We will work in the zero temperature limit for this EoS, which is characterized by two free parameters: the gluon condensate, $G_2$, and the large distance static ${\bar{q}}q$ potential, $V_1$ \cite{Mariani.etal:2017}. 

\subsection{Phase transition construction} \label{phasetransition}

As we mentioned before, there are two limiting approaches to study the phase transition between hadronic and quark matter: the Maxwell construction, where a sharp first-order phase transition with local charge conservation is considered, or the Gibbs construction, where a mixed phase exists and charge is conserved globally. The nature of the phase transition is related to the value of the surface tension in the hadron-quark interface \cite{Alford:2001zr,lugones:2013,ke:2014}. In this work we present results using the Maxwell construction. In addition, we simulate the Gibbs construction through a continuous interpolation between the hadronic and quark phases to mimic the mixing or percolation. Following previous  ideas, we use an interpolation function, $f_{\pm}(p)$, between the hadronic and quark phases given by
\begin{equation} \label{interpolate}
f_{\pm} (p) = \frac{1}{2}\left[1 \pm \tanh \left(\frac{p-p_{\rm trans}}{10 {\rm {b}}\,p_{\rm trans}} \right) \right],
  \end{equation}
\noindent  where + (-) is used to characterize the function used for pressures greater (smaller) than the transition pressure, $p_{\rm trans}$, and $\rm{b}$ is a free parameter that can be interpreted as a {\it mixing length} (note that when $\rm{b} \to 0$ we reproduce the Maxwell construction for the phase transition) \cite{Macher:2004vw,Masuda:2012ed,masuda:2013PTEP,Alvarez-Castillo2015,bai.et.al2018}. {{The resulting EoS reads
\begin{equation} \label{mixEOS}
\epsilon_{\rm MIX}(p) = \epsilon_{\rm H}(p)f_-(p)+ \epsilon_{\rm Q}(p)f_+(p),
  \end{equation}
\noindent where $\epsilon_{\rm H}(p)$ and $\epsilon_{\rm Q}(p)$ stands for the hadronic and quark EoS respectively.}}

Large values of ${\rm b}$ correspond to a broader mixed phase, typical for small values of the hadron-quark surface tension, $\sigma _{\rm HQ}$. To obtain a quantitative relationship between ${\rm b}$ and $\sigma _{\rm HQ}$, it is required to perform the ``full'' Gibbs construction for the mixed phase. The inclusion of interpolating function (\ref{interpolate}) allow us to analyze, in a qualitative way, if the nature of the phase transition could produce potentially observable effects on the frequencies of the oscillation modes. It is important to point out that in this work we can not relate the value of the parameter ${\rm b}$ with the hadron-quark surface tension, $\sigma _{\rm HQ}$. The study of finite-size effects related with the pasta-phase is out of the scope of this work.

To analyze the observational effects considering this heuristic construction of the mixed phase, we use three different values of $\rm{b}$: $\rm{b} = 0$ (Maxwell construction), $\rm{b} = 1$ and $\rm{b} = 3$.

In the left panel of Figure \ref{eos-mraio} we show a representative sample of the hybrid EoSes used in this work. In each case, results obtained using the three values of the ${\rm b}$ parameter are shown. In the right panel of Figure \ref{eos-mraio} we show the mass-radius curves obtained using this group of EoSes. In the case of Maxwell phase transition, the dot represents the first quark-hybrid star of the family with pure quark matter in its core. We remark that for the other two values of ${\rm b}$ the dot is not present because there are not quark-hybrid stars. The short names used to characterize the different quark EoSes in all the figures in this paper are explained in Tables \ref{short-names-fcm}, \ref{short-names-lnjl}, \ref{short-names-2sc+s} and \ref{short-names-nlnjl} where the values of the parameters selected for each case to attain the $2 M_\odot$ limit are detailed. {{In addition,the corresponding values of the CSS parameters are presented. The average of the square of the speed of sound, $\langle c_{\rm s}^2 \rangle$, is calculated for the stationary branch of quark-hybrid stars.}}

All the calculations regarding the construction of the mixed EoSes and the corresponding stellar structure have been computed using the {\it Neutron-Star-Object Research} (NeStOR) code \cite{Mariani.etal:2017}.

\section{The Cowling approximation} \label{cowling}

The study of relativistic stars oscillations has started with the pioneering work of Thorne and Campolattaro \cite{NS-oscillations}. The perturbation modes can be decomposed in two different families depending on the parity of the spherical harmonics that appear in their respective decomposition: odd (or axial), which produce toroidal deformations, and even, (or polar) which result in spheroidal deformations. The polar modes (to which we are devoted in this paper) are classified according to the most important restoring force acting on a fluid element when it gets displaced from its equilibrium position \cite{LRRkokkotas}.

The most important fluid modes related to the emission of GWs are the pressure ($p$), the fundamental ($f$) and the gravity ($g$) modes. It is known that $p$-modes have greater frequencies than the $g$-modes and these families are separated by the $f$-mode \cite{LRRkokkotas}. Within the relativistic Cowling approximation, the purely gravitational modes ($w$-modes) can not be studied and only the fluid modes presented above can be analyzed. It is important to state that in cold and non-rotating stars the $g$-modes are excited only when a discontinuous EoS is used to describe the matter inside a compact object \cite{Finn:1987}.

In the cases in which metric perturbations are negligible, the set of differential equations needed to solve in order to study oscillation modes is greatly simplified. This idea was first applied for Newtonian stars by Cowling \cite{cowling:1941} and several years after it was extended to the relativistic arena \cite{g-modes}. It is important to remark that in the relativistic Cowling approximation GWs are not emitted, because perturbations to the background metric are not taken into account. For this reason, as there is no damping, the frequencies of the modes are real numbers. {When considering spherically symmetric background spacetime characterized by a line element given by
  \begin{equation} \label{line_element}
{\rm d}s^2 = -e^{2\Phi (r)}{\rm d}t^2 + e^{2\Lambda (r)}{\rm d}r^2 + r^2 {\rm d}\theta ^2 + r^2 \sin ² \theta {\rm d}\phi ^2,
    \end{equation}
}
\noindent the equations needed to solve in order to find these frequencies are presented in \cite{sotani-cowling}

\begin{eqnarray} \label{eq:modes}
  \frac{{\rm d}W (r)}{{\rm d}r} &=& \frac{{\rm d} \epsilon}{{\rm d}P} \left[\omega ^2 r^2 {\rm e}^{\Lambda (r) - 2 \Phi (r)} V(r) + \frac{{\rm d}\Phi (r)}{{\rm d}r} W(r)\right] - \ell (\ell + 1){\rm e}^{\Lambda (r) } V(r), \\
  \frac{{\rm d}V(r)}{{\rm d}r} &=& 2 \frac{{\rm d}\Phi (r)}{{\rm d}r} V(r) - \frac{1}{r^2}{\rm e}^{\Lambda (r) } W(r) . \nonumber
  \end{eqnarray}

\noindent The functions $V(r)$ and $W(r)$, along with the frequency $\omega$, characterize the Lagrangian perturbation vector associated with the fluid,
{
\begin{equation} \label{pert}
\xi ^i = \left(e^{-\Lambda (r)}W(r), -V(r)\partial _\theta, -V(r) \sin ^{-2} \theta \partial _\phi \right)r ^{-2} Y_{\ell m}(\theta , \phi),
\end{equation}
}
\noindent where $Y_{\ell m}(\theta , \phi)$ is the $\ell m$-spherical harmonic. We solve (\ref{eq:modes}) on a fixed background metric from the origin $(r \sim 0)$, where the solutions behave approximately like

\begin{equation} \label{ini}
W(r) \sim Ar^{\ell +1}, \qquad V(r) \sim - \frac{A}{\ell}r^\ell ,
\end{equation}
where $A$ is an arbitrary constant. The other boundary condition that needs to be fulfilled is that the Lagrangian perturbation to the pressure, $\Delta P$, must vanish at the star's surface ($r = R$). Such condition reads

\begin{equation} \label{BCond}
\omega ^2 {\rm e}^{\Lambda (R) - 2 \Phi (R)} V(R) + \frac{1}{R^2}\frac{{\rm d}\Phi (r)}{{\rm d}r}\bigg|_{r = R} W(R) = 0.
\end{equation}
The previous equations can be generalized to the case in which the EoS has a discontinuity at $r = r_{\rm t}$. The following additional connection formula{, which are the continuos condition for $W$ and $\Delta P$ (the functions that appear in the linearized perturbations of Einstein equations of General Relativity within the Cowling approximation), re-written in terms of the functions $W$ and $V$ which turns out to be discontinuous,} must be fulfilled \cite{sotani-full}

\begin{eqnarray} \label{connection-formula}
  W_+ &=& W_-, \nonumber \\ 
  V_+ &=& \frac{{\rm e}^{2\Phi}}{\omega ^2 r^2_{\rm t}} {\rm e}^{-\Lambda}\times \left(\frac{\epsilon _- + P}{\epsilon_+ -P}\left[\omega ^2 r^2_{\rm t}{\rm e}^{\Lambda -2\Phi}V_- + \Phi ^{\prime} W_-\right] - \Phi ^{\prime} W_+\right), 
  \end{eqnarray}
where the minus (plus) subindex corresponds to quantities before (after) the phase transition.

\begin{figure}[h]
  \centering
  \includegraphics[width=0.34\textwidth, angle = -90]{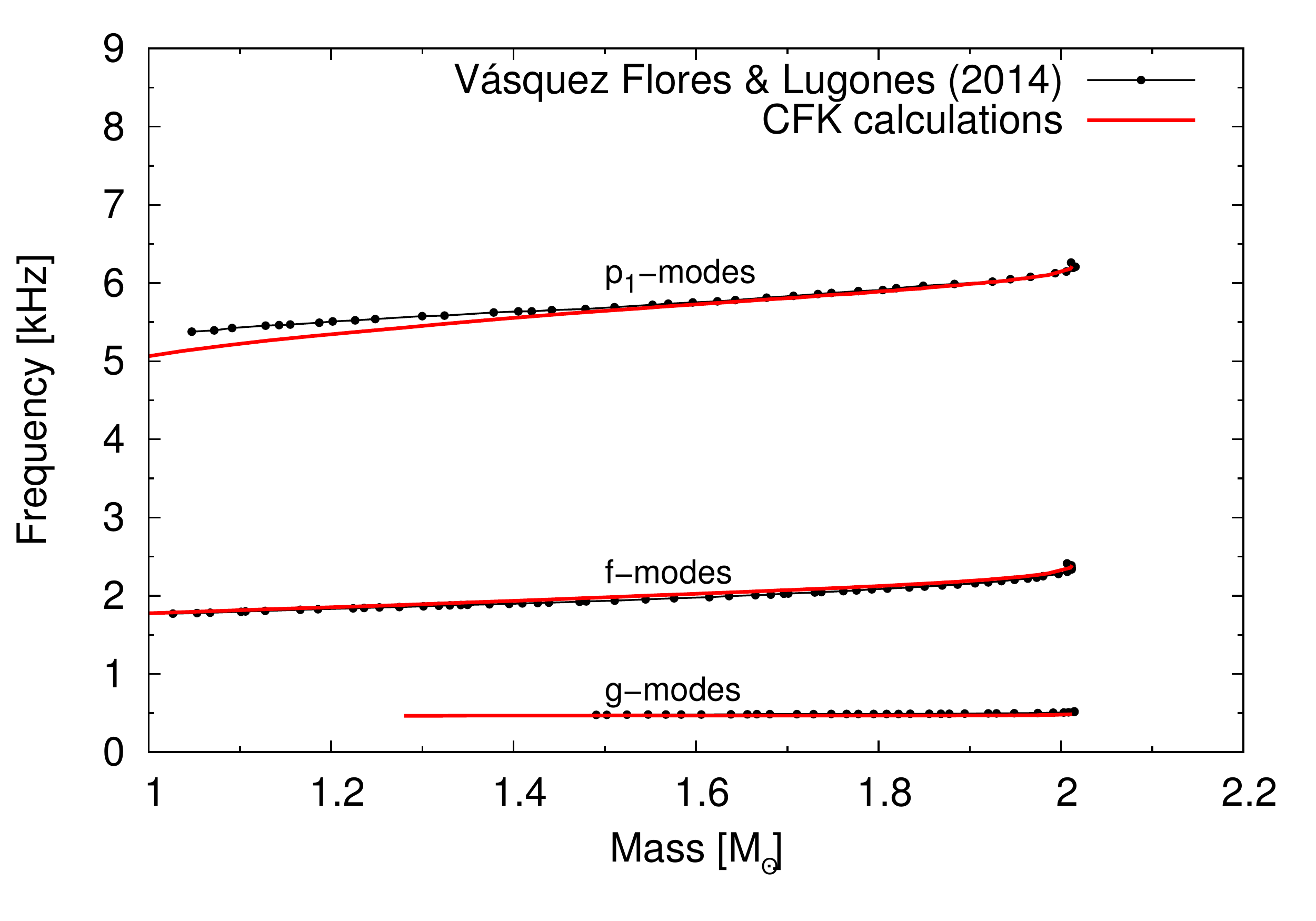}
  \caption{(Color online) For hybrid stars constructed using GM1 hadronic EoS and a modified bag with parameters $B_b = 59.2$MeVfm${}^{-3}$ and $a_4 = 0.55$ for the quark phase, we present with lines our calculations and with dots results taken from  V\'asquez-Flores and Lugones \cite{vasquez-lugones:2013}. Slight differences are seen only for the $p_1$ mode at low masses.
}
  \label{comp-lugones}
\end{figure}

It is important to note that for neutron stars with masses above $\sim 1 M_\odot$, the frequencies obtained using the relativistic Cowling approximation differ by 10-30\%, for the $f$-mode, when compared with those based on calculations considering the complete linearized equations of General Relativity \cite{Finn:1988,accuracy-cowling,vasquez-lugones:2013,chirenti:2015,vasquez-lugones:2017,vasquez-flores:2017}. Moreover, it has been shown that the errors decrease when the stellar compactness increases \cite{chirenti:2015}. Thus, for the most massive stars obtained using our EoSes, the errors in the $f$-mode frequencies are not greater than $\sim$ 15\%. This discrepancy is less than $\sim 10$ \% for the first $p$-modes and less than $\sim 5$ \% for the $g$-modes \cite{sotani-full,accuracy-cowling}. Therefore, the results obtained using this method are both qualitatively and quantitative good approximations for the $p$ and $g$-modes and only a qualitatively good approximation for the $f$-modes.

\subsection{The Numerical Method} \label{numerical}

In this subsection, we present a brief description of the numerical procedure in the CFK code used to obtain the main results of this work.

The numerical integration of (\ref{eq:modes}) with initial conditions (\ref{ini}) was performed using a Runge-Kutta-Fehlber (RKF) predictor-corrector integration method \cite{numerical-recipies}. The RKF value of the integration step was chosen in order to have several integration steps between any two layers of the stellar structure.

To solve the boundary conditions (\ref{BCond}) for continuous EoS we use a Newton-Raphson algorithm coupled with Ridders' method to enhance convergence and avoid numerical instabilities in the equations to solve. In the case of discontinuous EoS, the integration also starts at the center of the star using (\ref{ini}) and is performed up to $r = r_ {\rm t}$, where connection equations (\ref{connection-formula}) are used to define new initial conditions to continue the numerical integration until the surface of the star, where condition (\ref{BCond}) is fulfilled using the procedure explained before.

CKF code was developed to calculate the frequencies of a given mode for the complete family of stable stars constructed with a given EoS. An initial seed for the frequency, $\omega^2_0$, is defined for the star with the lowest mass to find the true frequency which is used as the initial seed for the next model. This is performed until the oscillation frequency for the last stable star of the family is obtained.

To test the CFK code we have reproduced some of the results of V\'asquez-Flores and Lugones \cite{vasquez-lugones:2013}{{, in particular the frequencies of the $g$, $f$ and $p_1$-modes obtained, using the Cowling approximation}}. As an example, in Figure \ref{comp-lugones}, we show the frequencies obtained for $g$, $f$ and $p_1$-modes considering hybrid stars constructed with the hadronic GM1 parameterization and a modified bag model for the quark phase with $B_b = 59.2$MeVfm${}^{-3}$ and $a_4 = 0.55$ \cite{vasquez-lugones:2013}. For the external crust we use the classical BPS EoS. The agreement between the results obtained using the CFK code and previous results shows the robustness and reliability of our code. {{The slight discrepancies seen for low mass stars (especially for the $p_1$-modes) are a consequence of differences in the EoS used to describe matter in the crust of the compact object.}}

\section{Results} \label{results}

We have obtained the frequencies of the $g$, $f$ and $p_1$-modes of both quark-hybrid and hybrid stars constructed using the EoSes described in Section \ref{EoSes}.

\begin{figure*}[h]
  \centering
  \includegraphics[width=0.7\textwidth, angle = -90]{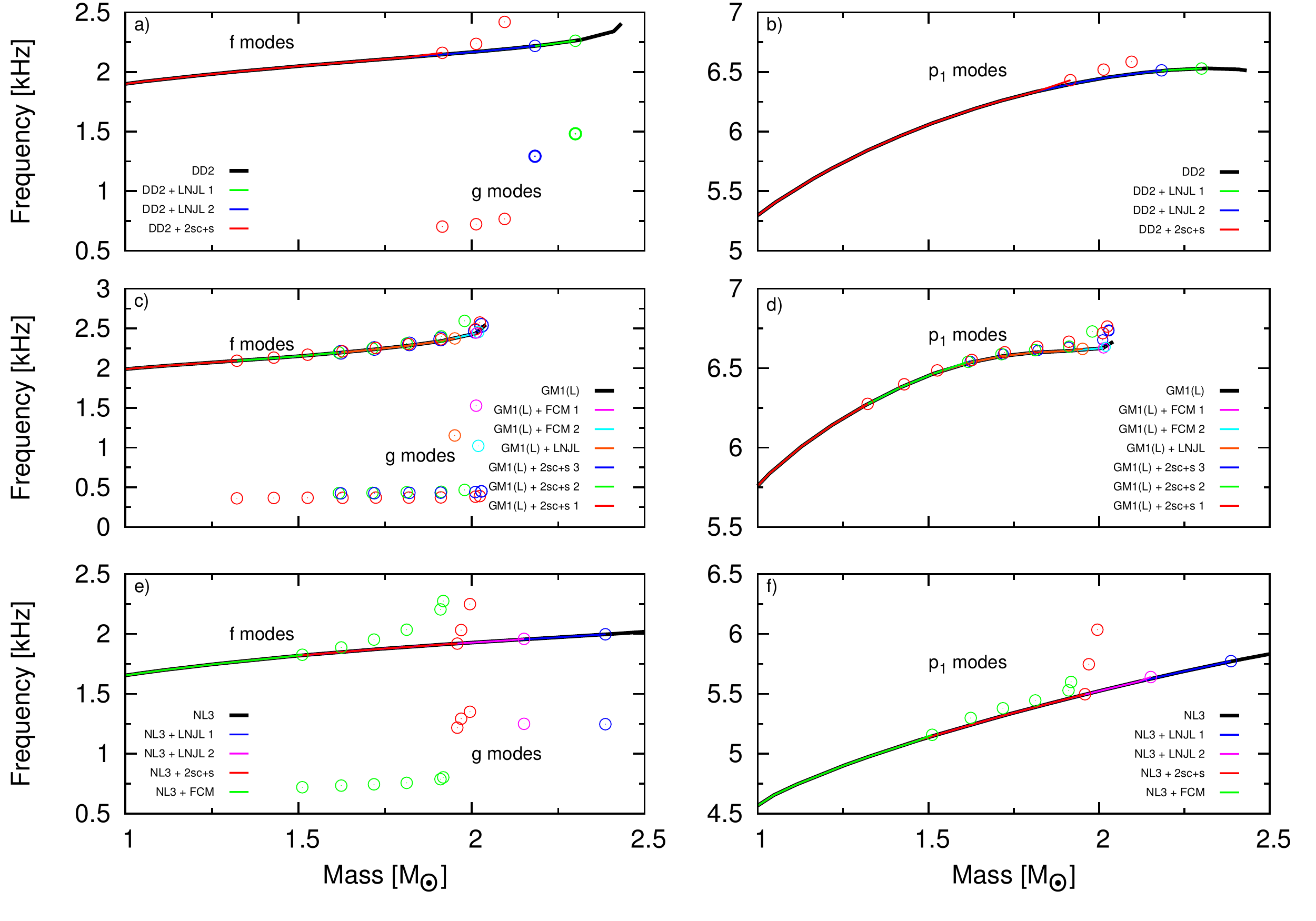}
  \caption{(Color online) Frequencies of $g$, $f$ and $p_1$-modes for different hybrid EoSes. With lines we show the calculated frequencies of purely hadronic stars and with open circles the ones for stars with pure quark matter in their cores. For details regarding the EoSes see Table \ref{short-names-fcm}, Table \ref{short-names-lnjl} and Table \ref{short-names-2sc+s}.
}
  \label{multiplot-frec2}
\end{figure*}

In Figure \ref{multiplot-frec2} we present a series of plots showing the frequencies of the $g$, $f$ and $p_1$-modes considering quadrupolar perturbations ($\ell = 2$) as a function of the compact object's mass for different EoSes. The frequencies are shown with lines for purely hadronic stars and with open circles for stars with pure quark matter in their cores. The different colors indicate the different EoSes used in panels (a) and (b); (c) and (d) and (e) and (f). We found that for some of our EoSes, only an extremely short branch of connected quark-hybrid stars exists ($\Delta M \sim 10^{-2} - 10^{-3} M_\odot $). For this reason only the $g$-mode frequency corresponding to the last stable star of this family is plotted.

For different hybrid EoSes constructed using the hadronic DD2 parametrization, we show in panel (a) the frequencies of the $g$ and $f$-modes as functions of the stellar mass. For comparison, in black we show the results obtained with the purely hadronic EoS. We can see that the appearance of quark matter in the core of a compact object produces a decrease in the mass of the most massive star. This effect is more evident when color superconductivity is taken into account. In this case, the most massive quark-hybrid star is $\sim 0.3M_\odot$ less massive than the most massive hadronic star. For the other cases analyzed (without color superconductivity) this effect is less noticeable. Another effect of color superconductivity is to produce an increase in the frequencies of the $f$-modes. For the most massive quark-hybrid star this increase is $\sim 25$ \%. In panel (b) we show the results for the $p_1$-mode, which are qualitatively similar, but the relative increase in the frequencies is smaller than for the $f$-modes ($\sim 5$ \% for the most massive star with color superconductivity).

In panels (c) and (d), we show the same results as in panels (a) and (b), but using the hadronic GM1(L) EoS. The qualitative results are similar to those obtained using the DD2 hadronic EoS. However, the differences between the frequencies of the $f$-mode in stars with pure quark matter cores as well as the decrease in the mass of the most massive quark-hybrid stellar model are less noticeable. In the case of the $p_1$-mode, the effects are qualitatively similar, but there is a degeneracy (in the sense that stars constructed with the same quark EoS but different sets of parameters have the same $p_1$ oscillation frequency for a given mass) that could be broken if precise observations of $f$ and $g$-modes were available for high-mass compact objects.

In panels (e) and (f) the hadronic EoS used is the NL3 parametrization. Beside the onset of the $g$-modes, it can be seen that the appearance of quark matter in the cores of these quark-hybrid stars produces an increase (for a fixed mass) of the $f$ and $p_1$-modes frequencies. In the case of the $f$-mode, both the FCM and the 2SC+s quark EoSes produce a significant increase in the frequencies. For the most massive quark-hybrid star (with a mass $0.6M_\odot$ smaller than the most massive hadronic star) such increase is $\sim 30$\%. On the contrary, in the case of the $p_1$-mode, the change in the frequency for the quark-hybrid EoS constructed with the FMC is not as noticeable ($\sim 3$ \% for the most massive star) as in the case in which color superconductivity is considered ($\sim 10$ \% for the most massive star).

\begin{figure}[h]
  \centering
  \includegraphics[width=0.34\textwidth,angle = -90]{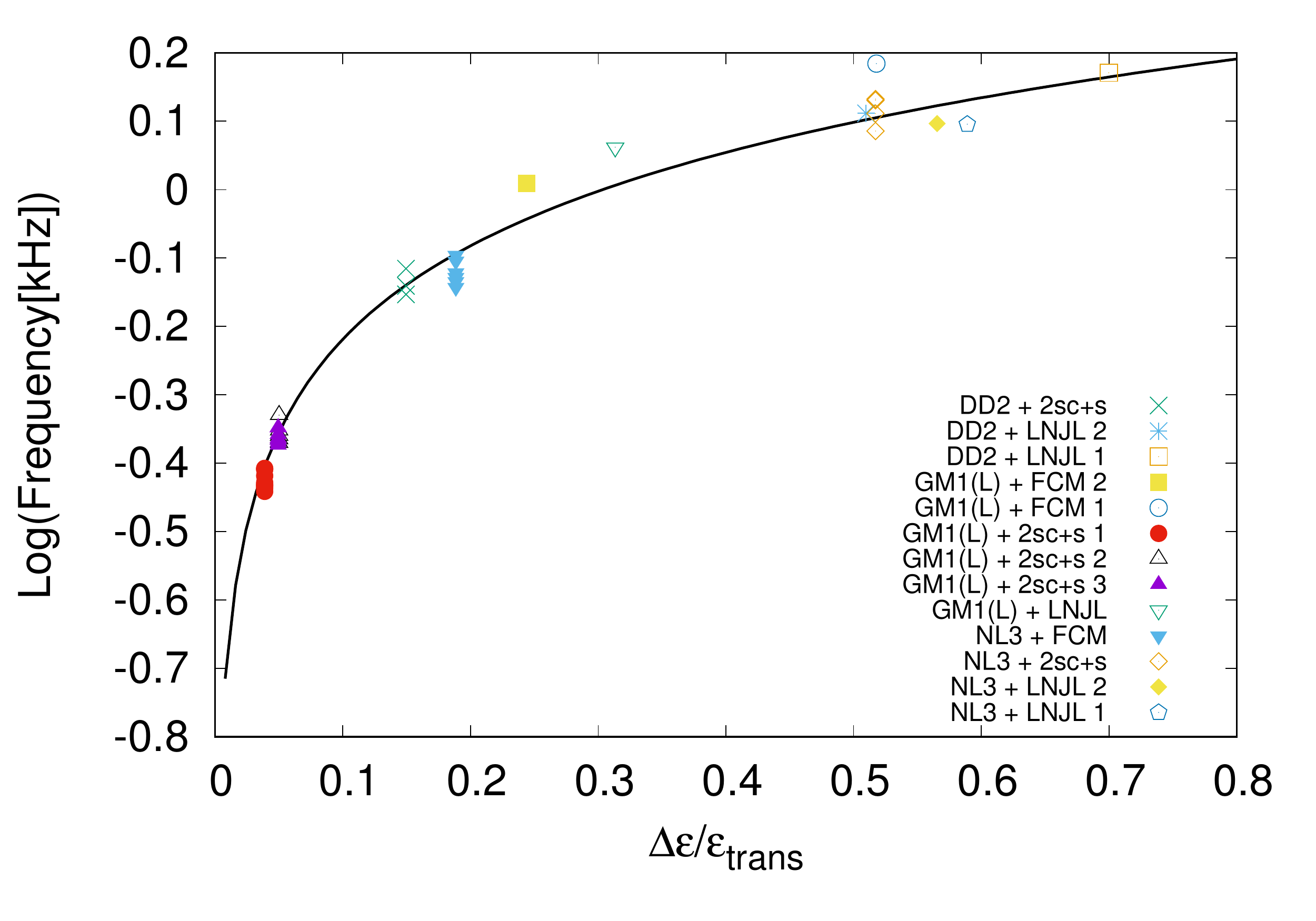}
  \caption{(Color online) Relationship between the decimal logarithm of the frequency of the $g$-mode and the CSS parameter ${\rm x}_{\rm CSS} \equiv \Delta \epsilon / \epsilon _{\rm trans}$. A fit to the data is presented with a black line.}
  \label{rel-delta-e-g}
\end{figure}

Summarizing, the detection of GWs from isolated high-mass neutron stars could be a powerful tool to discern between different EoSes compatible with the $2M_\odot$ limit imposed by observations. It can be seen that for a given stellar mass, for stars constructed using GM1(L) EoS, the frequencies of $f$ and $p_1$-modes
are greater than for those constructed with DD2 EoS. For stars constructed with the NL3 EoS, the frequencies are smaller. This situation changes slightly for high-mass quark-hybrid stars (see, for example, panel (d) of Figure \ref{multiplot-frec2}. This possible degeneration shown in Figure \ref{multiplot-frec2}, panel (d), would be broken if $g$-modes are detected.

\begin{figure}[h]
  \centering
  \includegraphics[width=0.34\textwidth,angle = -90]{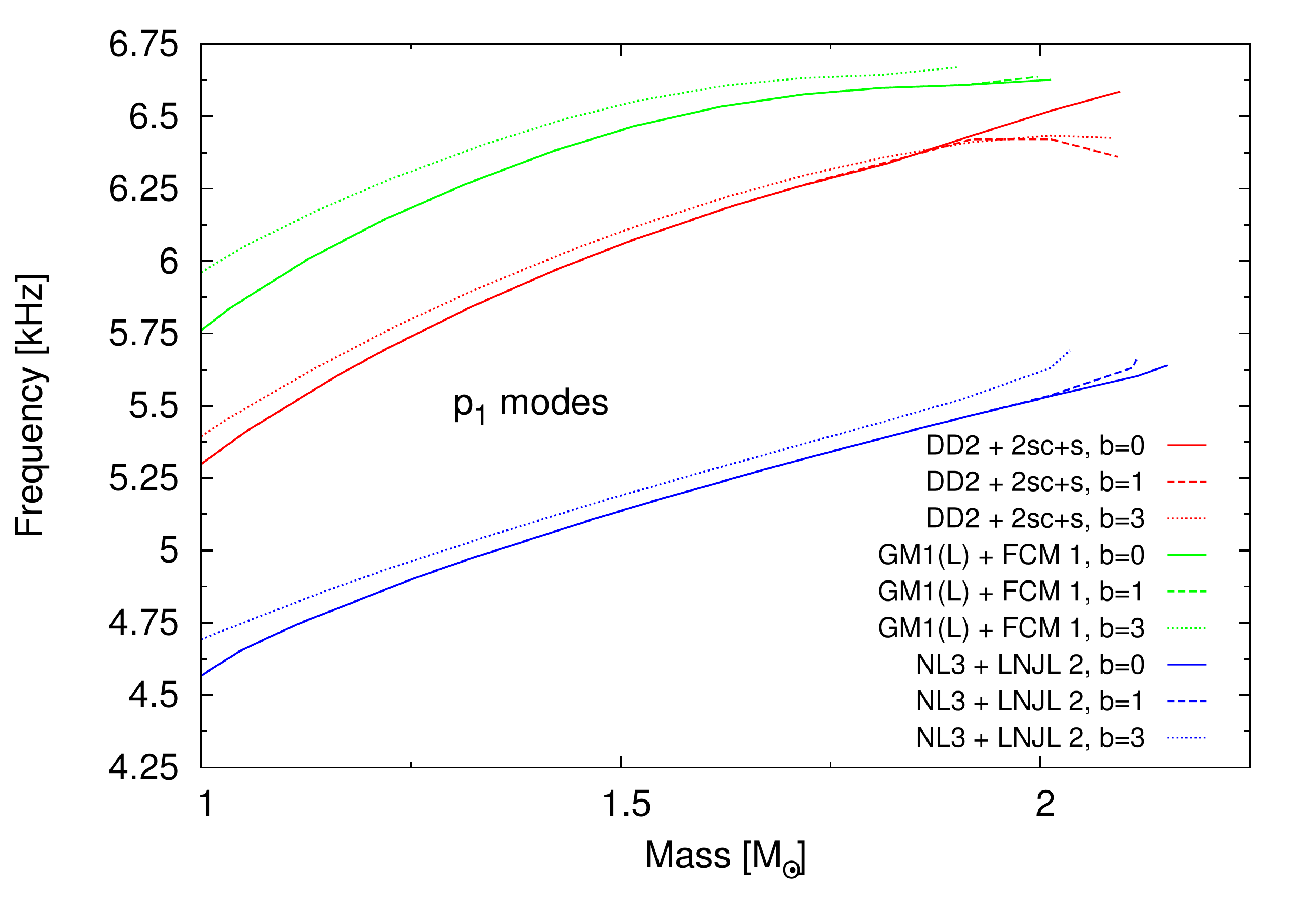}
  \includegraphics[width=0.34\textwidth,angle = -90]{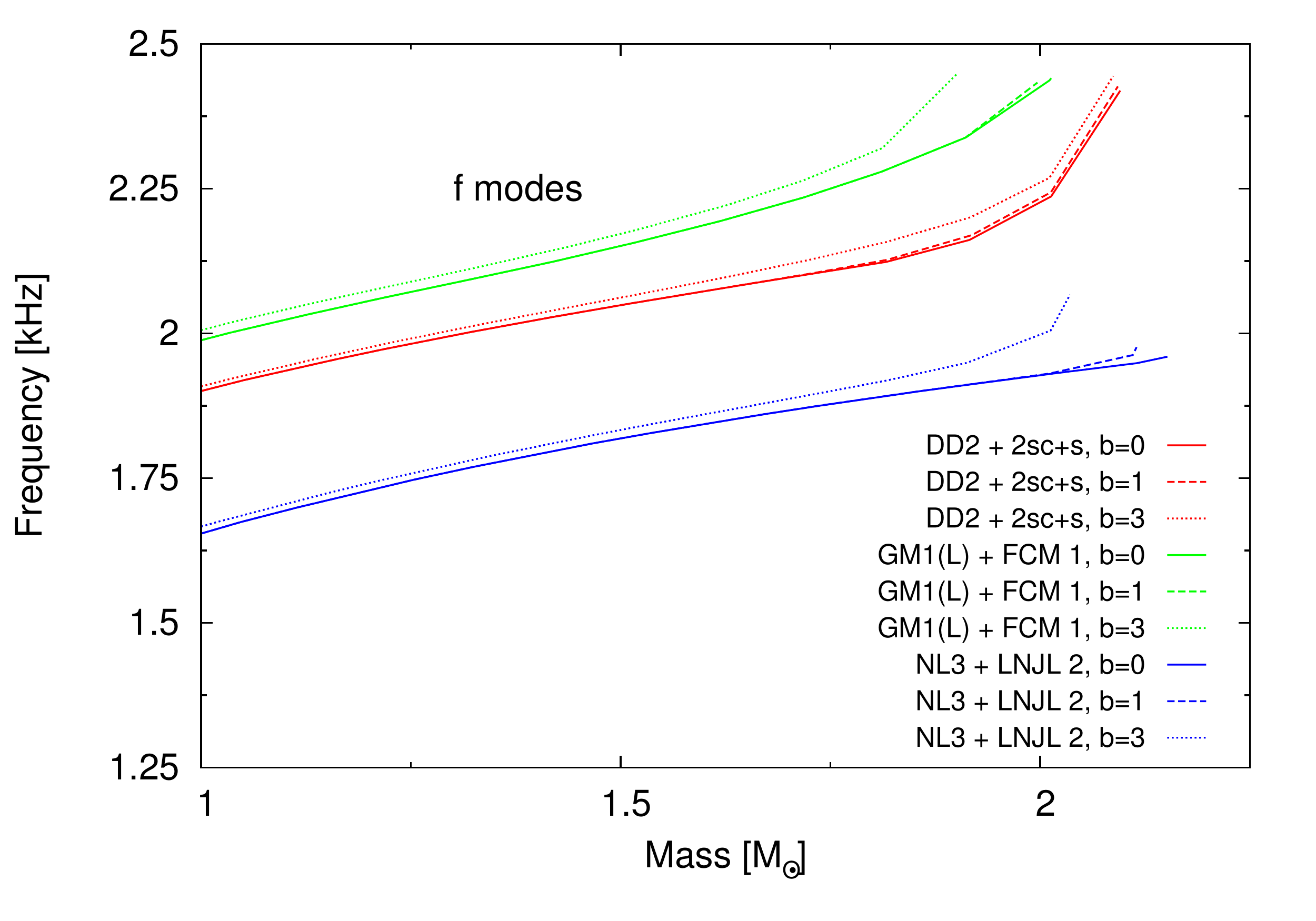}
  \caption{(Color online) Frequencies of the $p_1$ (left) and $f$-modes (right) for the EoSes presented in Figure \ref{eos-mraio} for different {\it mixing length} parameter values. The effect of increasing the value of this parameter is a rise in the frequencies of the $p_1$-mode for low mass stars and the same for the $f$-mode but for high-mass stars. These effects are more evident for ${\rm b} = 3$.}
  \label{effect-of-b}
\end{figure}

Another important result of our work is the correlation of the decimal logarithm of the $g$-mode frequencies and the quotient between the quark-hadron energy gap and energy density at the transition, ${\rm x}_{\rm CSS} \equiv \Delta \epsilon / \epsilon _{\rm trans}$, one of the parameters of the constant speed of sound (CSS) parameterization for quark matter \cite{CSS-original}.

\begin{table}[h]
\centering
\caption{Short names for hybrid EoSes constructed with the FCM EoS for the quark phase. The parameters of the FCM are shown in each table row. {{We also present the associated values of the CSS parametrization for each model.}}}
\label{short-names-fcm}
{
\begin{tabular}{c c c c c c}
\hline
\multirow{2}{*}{\begin{tabular}[c]{@{}c@{}}Short\\ Name\end{tabular}} & \multicolumn{2}{c}{FCM parameters}  & \multicolumn{3}{c}{CSS parameters} \\ \cline{2-6} 
                                                                      &  {$V_1$ {[}GeV{]}} & $G_2$ {[}GeV${}^4${]} & $p_{\rm trans} / \epsilon _{\rm trans}$ & $\Delta \epsilon / \epsilon _{\rm trans}$ & $\langle c_{\rm s}^2 \rangle$ \\ \hline
GM1(L) + FCM 1                                                        & 0.10           & 0.006      & 0.271 & 0.518 &  0.295       \\ \hline
GM1(L) + FCM 2                                                        & 0.15           & 0.002    & 0.275 & 0.244 &  0.279          \\ \hline
NL3 + FCM                                                             & 0.15           & 0.002   & 0.118 & 0.189 &  0.268         \\ \hline
\end{tabular}}
\end{table}

In Figure \ref{rel-delta-e-g} we show the frequencies of the $g$-modes obtained for the different hybrid EoSes as a function of ${\rm x}_{\rm CSS}$. Moreover, the fitting function, $y_{\rm fit} = c_1 \log ({\rm x}_{\rm CSS})+ c_2$, is also indicated with black solid line. The fitted parameters are $c_1= 0.454 \pm 0.031$ and $c_2 = 0.235 \pm 0.023$. The root mean square of residuals for the fit is $R^{\rm rms} = 0.048$ showing a strong correlation between these quantities. A correlation between the $g$-modes frequencies and the parameter ${\rm x}_{\rm CSS}$ could be expected because these modes are suppressed in continuous EoSes. However, the fact that such a good fit is obtained using a simple function is an interesting result. The wide range of hybrid EoSes used in this work, suggests that the obtained relationship is of {\it universal} nature. Such relationship could be used to better understand the behavior of matter inside compact objects. After a detection of a $g$-mode, it could be possible to infer the existence of a sharp phase transition inside the star and also deduce a value for one of the CSS parameters (${\rm x}_{\rm CSS}$). This information would be used to constrain the hybrid EoSes. {{Other relationships between NS observables and physical parameters can be found in the literature, for example, the upper limit on the mass and rotation speed set by causality \cite{1997ApJ...488..799K} or the value of the fiducial energy density up to which the modern neutron star matter EoSes are believe to be reliable \cite{1996ApJ...470L..61K}. Another interesting relationship has recently been proposed between the maximum mass of a compact object and the maximum speed of sound of matter in the inner core of neutron stars \cite{2017PhRvC..95b5802S}.}}

\begin{table}[h]
\centering
\caption{Short names for hybrid EoSes constructed with the LNJL EoS for the quark phase. The vector interaction parameter, $\eta_{\rm v}$, used in the LNJL EoS is shown in each table row. {{Moreover, we present the associated values of the CSS parametrization for each case.}}}
\label{short-names-lnjl}
{
\begin{tabular}{c c c c c c}
\hline
\multirow{2}{*}{\begin{tabular}[c]{@{}c@{}}Short\\ Name\end{tabular}} & \multicolumn{2}{c}{LNJL parameter}  & \multicolumn{3}{c}{CSS parameters} \\ \cline{2-6} 
                                                                      & \multicolumn{2}{c}{$\eta_{\rm v}$} & $p_{\rm trans} / \epsilon _{\rm trans}$ & $\Delta \epsilon / \epsilon _{\rm trans}$ & $\langle c_{\rm s}^2 \rangle$    \\ \hline
DD2 + LNJL 1                                                          & \multicolumn{2}{c}{0.30}   & 0.322 & 0.699 & 0.230         \\ \hline
DD2 + LNJL 2                                                          & \multicolumn{2}{c}{0.15}    & 0.274 & 0.509 & 0.204        \\ \hline
GM1(L) + LNJL                                                         & \multicolumn{2}{c}{0.00}   & 0.241 & 0.313 & 0.179           \\ \hline
NL3 + LNJL 1                                                          & \multicolumn{2}{c}{0.30}   & 0.271 & 0.589 & 0.244        \\ \hline
NL3 + LNJL 2                                                          & \multicolumn{2}{c}{0.15}    & 0.217 & 0.565 & 0.350         \\ \hline
\end{tabular}}
\end{table}

If the existence of a mixed phase is considered using our heuristic model, families of compact objects with lower maximum masses are obtained. This effect is enhanced for larger values of the $\rm{b}$ parameter. Regarding the oscillation modes, the mixed phase inhibits the excitation of $g$-modes since the resulting interpolated EoS do not present discontinuities. The general result for the $f$-mode is that their frequencies increase with the ${\rm b}$ value. For stars with masses greater than $\sim 1.8M_\odot$, the changes in the frequencies for different values of the $\rm{b}$ parameter become significant ($\sim 10$ \% for the most massive hybrid star constructed using ${\rm b} = 3$). For $\rm{b} = 1$ this effect is not relevant for any compact-object mass.

In Figure \ref{effect-of-b}, we show these effects for representative cases (see caption for details). In the right panel, we show the effect on the $f$-modes, comparing the results obtained for phase transitions constructed using Maxwell formalism (${\rm b} = 0$) with the ones obtained using $\rm{b}=1$ and $\rm{b} = 3$. In the left panel we do the same comparison for the $p_1$-mode. A particular feature is observed when color superconductivity is considered. Neutron stars with masses grater than $\sim 1.8M_\odot$ present smaller $p_1$-mode frequencies than the ones obtained using Maxwell construction (for the case in which ${\rm b} = 3$, the difference is $\sim 4$ \% for the most massive star).

\begin{table}
\centering
\caption{Short names for hybrid EoSes constructed with the 2SC+s EoS for the quark phase. The parameters of the 2SC+s are shown in each table row. {{Moreover, we present the associated values of the CSS parametrization for each case.}}}
\label{short-names-2sc+s}
{
\begin{tabular}{c c c c c c c c}
\hline
\multirow{2}{*}{\begin{tabular}[c]{@{}c@{}}Short\\ Name\end{tabular}} & \multicolumn{4}{c}{2SC+s parameters}     & \multicolumn{3}{c}{CSS parameters}                                                   \\ \cline{2-8} 
                                                                      & $M_{\rm s}^0$ {[}MeV{]} & $\eta _{\rm qq}$ & $\eta _{\rm v}$ & $\Delta _{\rm s}$ {[}MeV{]} & $p_{\rm trans} / \epsilon _{\rm trans}$ & $\Delta \epsilon / \epsilon _{\rm trans}$ & $\langle c_{\rm s}^2 \rangle$  \\ \hline
DD2 + 2SC+s                                                           & 500                     & 1.20             & 1.00            & 50      & 0.185 & 0.150 & 0.398                    \\ \hline
GM1(L) + 2SC+s 1                                                      & 500                     & 1.20             & 1.00            & 50        & 0.117 & 0.039 & 0.390                    \\ \hline
GM1(L) + 2SC+s 2                                                      & 500                     & 1.20             & 0.90            & 50       & 0.148 & 0.050 & 0.390                     \\ \hline
GM1(L) + 2SC+s 3                                                      & 600                     & 1.20             & 0.90            & 50      & 0.148 & 0.049 & 0.420                      \\ \hline
NL3 + 2SC+s                                                           & 600                     & 1.10             & 0.85            & 10       & 0.186 &  0.521 & 0.400                     \\ \hline
\end{tabular}}
\end{table}

\begin{table}
\centering
\caption{Short names for hybrid EoSes constructed with the NLNJL EoS for the quark phase. The vector interaction parameter, $\eta_{\rm v}$, used in the NLNJL EoS is shown in each table row. {{Moreover, we present the associated values of the CSS parametrization for each case.}} }
\label{short-names-nlnjl}
{
\begin{tabular}{c c c c c c}
\hline
\multirow{2}{*}{\begin{tabular}[c]{@{}c@{}}Short\\ Name\end{tabular}} & \multicolumn{2}{c}{NLNJL parameter} & \multicolumn{3}{c}{CSS parameters}              \\ \cline{2-6} 
                                                                      & \multicolumn{2}{c}{$\eta_{\rm v}$}  & $p_{\rm trans} / \epsilon _{\rm trans}$ & $\Delta \epsilon / \epsilon _{\rm trans}$ & $\langle c_{\rm s}^2 \rangle$    \\ \hline
DD2 + NLNJL 1                                                          & \multicolumn{2}{c}{0.00}    & 0.252 & 1.452 & 0.261        \\ \hline
DD2 + NLNJL 2                                                          & \multicolumn{2}{c}{0.09}  & 0.341 & 0.981 & 0.448           \\ \hline
\end{tabular}}
\end{table}

\begin{figure}[h]
  \centering
  \includegraphics[width=0.34\textwidth,angle = -90]{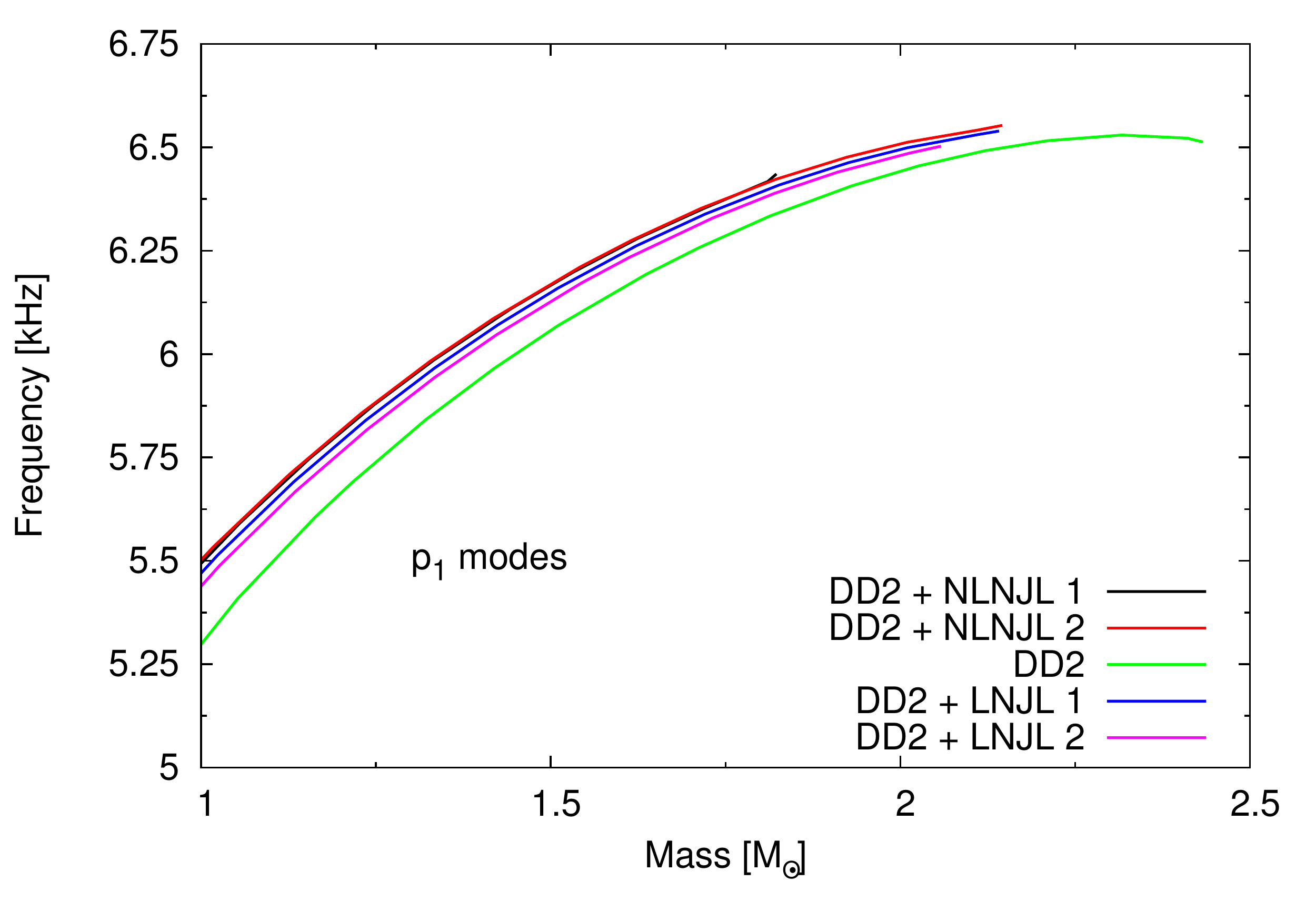}
  \includegraphics[width=0.34\textwidth,angle = -90]{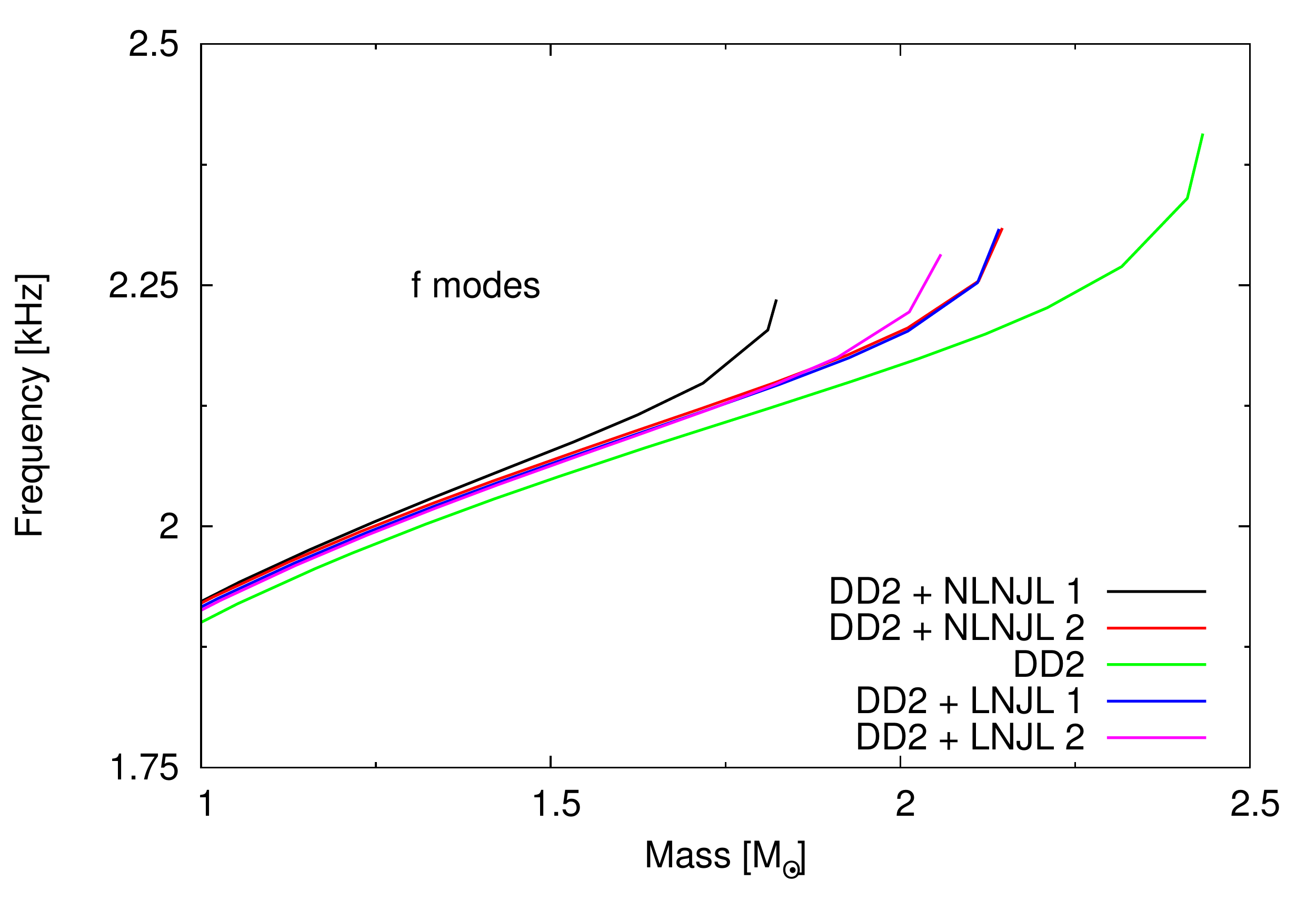}
  \caption{(Color online) Comparison between $p_1$ and $f$-modes of hybrid stars constructed with LNJL and NLNJL models and the interpolating function with a {\it mixing length} set to ${\rm b} = 3$. The hadronic parametrization used is the DD2 (see Table \ref{short-names-lnjl} and Table \ref{short-names-nlnjl} for details of the free parameters of the quark EoSes).}
  \label{loc-vs-nonloc}
\end{figure}

We have also performed calculations using the NLNJL model. For this quark EoS, it has been shown, using the NL3 hadronic EoS, that it is impossible to construct quark-hybrid stars because the appearance of quark matter in their inner core destabilizes the star \cite{rs-etal:2016}. We found the same behavior for the other two hadronic EoSes used in this paper. This suggests that the detection of $g$-modes might be used to discard NLNJL as a possible EoS to describe quark matter if a sharp phase transition inside hybrid stars exist. When the existence of a mixed phase is considered, our calculations show that very accurate observations of the oscillation modes would be needed to distinguish between the LNJL and NLNJL models because only slight changes in the frequencies of $f$ and $p_1$-modes are observed (see Figure \ref{loc-vs-nonloc}, as an illustration of this general result).

\section{Summary and Discussion} \label{discussion}

NeStOR working alongside with CFK allow us to study and analyze quadrupolar stellar oscillations of hybrid compact stars within the relativistic Cowling approximation.

We have used three different EoSes to describe the hadronic phase: the relativistic mean-field model using the NL3 classical parametrization, the DD2 parametrization, in which density-dependent coupling constants are included, and a modified GM1 parameterization in which a density-dependent isovector meson-baryon coupling constant is included. For the description of the quark phase, several EoSes have been used: LNJL, NLNJL and FCM effective models. In the LNJL case, color superconductivity in the 2SC+s phase has been included. To construct the phase transition, we have used the Maxwell construction and a heuristic approach to study the mixed phase in which hadrons and quarks co-exists. In all cases, the free parameters of these EoSes combination have been chosen to be consistent with the observed $2 M_\odot$ mass limit for neutron stars. 

We have found the frequencies of the $g$, $f$ and $p_1$-modes for {compact stars constructed using 15 cold dense matter EoSes with sharp hadron-quark phase transition. Moreover, we have also studied the appearance of a mixed phase in these EoSes and its effect on such frequencies.} The three families of modes could be easily differentiated from an observational point of view. Although their frequencies are sensitive to differences in the hybrid stars composition, there exist an overlapping between the curves for different EoSes, specially in that cases of the fundamental and first pressure modes for stars with masses below $1.8M_\odot$. The fact that this degeneracy is not present in the $g$-modes, makes them a tool that could shed some light into determining the nature of the NSs EoS.

Using the CFK code, we have obtained the frequencies of the $g$-modes produced by the existence of a sharp discontinuity in some of our EoSes. The values for these frequencies are between $\sim 0.5$ kHz and $\sim 1.5$ kHz. It is important to note that there are several physical phenomena (thermal effects, chemical gradients in the outer part of compact stars, stellar rotation or magnetic fields) that could produce an overlap with the $g$-modes that we have found. However, the frequencies of $g$-modes associated with thermal effects, chemical gradients in the outer part of compact stars or $f$-modes modified by rotation or magnetic fields have frequencies which are systematically lower than $1$ kHz \cite{low-frec-g-modes,sotani:2007,kantor:2014,dommes:2016,lugones-qs:2016}. For this reason, we suggest that future observations of $g$-mode GW emission in the range $1$ kHz -- $1.5$ kHz could indicate{, if simultaneous detection of the $f$-mode is available,}  the presence of pure quark matter in the inner core of neutron stars, that is, an evidence of the existence of quark-hybrid stars. 

For stars with masses above $\sim 1M_\odot$, the effects of a mixed phase are not evident for $p_1$-modes. In addition, in the case of the fundamental mode, for compact stars with masses above $\sim 1.8M_\odot$, minor changes appear (for most of the EoSes used in this work). These facts suggest that the detection of oscillation modes would not be a useful tool to draw any conclusion about the breadth of the mixed phase. On the other hand, when color superconductivity is considered, the frequencies of the $p_1$-modes decrease for compact stars with masses above $\sim 1.8 M_\odot$. It may be interesting to analyze how our results are affected when the mixed phase is constructed using the full Gibbs formalism instead of the heuristic model used in this work.

The results obtained for the behavior of the $g$-modes frequency as a function of the $x_{\rm CSS}$ parameter allow us to highlight the virtues of GW analysis as an important tool to understand the internal composition of compact objects. We found a very good matching in the fitting between the logarithm of the frequency of this mode and $x_{\rm CSS}$. Although some correlation between this two quantities was expected to exist, since the $g$-modes are only present when an energy gap, $\Delta \epsilon$, in the hybrid EoS occurs, the result obtained is promising. This correlation could be used to better understand the behavior of matter inside compact objects. 

Pulsating compact objects might produce a rich spectrum of GWs beyond the frequency modes studied in this paper. Although the EoSes used in this work cover a broad range of modern theoretical EoSes able to reproduce the $2M_\odot$ observations, the conclusions drawn in this work might have to be revised if compact stars constructed with other EoSes emit GWs in the same frequencies than the obtained in our calculations. Nevertheless, the conclusion regarding the fact that $g$-mode GW emission with frequencies in the range of $1.0 - 1.5 \rm{kHz}$ should be interpreted as emitting hybrid stars with pure quark matter in their cores and the presence of a sharp phase transition between quark and hadronic matter, should remain valid.

\bibliographystyle{JHEP}
\bibliography{IFRS}

\acknowledgments
The authors want to thank Prof. Dr. G. Lugones for helpful comments during the early stages of this project, to Prof. Dr. H. Vucetich for pointing out the virtues of Ridder's method. We would also like to thank the the Editors of JCAP and the anonymous referee for their time and her/his suggestions that helped improve substantially the first version of the paper. IFR-S, MM and MGO acknowledge support from Universidad Nacional de La Plata and CONICET under Grants G140, G157 and PIP-0714. OMG thanks CONICET and Universidad Nacional de La Plata for financial support under Grants PIP-0436 and G144.


\end{document}